# Van der Waals stacked multilayer in-plane graphene/hexagonal boron nitride heterostructure: its interfacial thermal transport properties

Ting Liang, Ping Zhang[*], Man Zhou, Peng Yuan and Daoguo Yang

School of Mechanical and Electrical Engineering, Guilin University of Electronic Technology

No. 1 Jinji Road, Guilin, Guangxi 541004, China

*corresponding author, E-mail: *pingzhang@guet.edu.cn*

## Abstract

Combining both vertical and in-plane two-dimensional (2D) heterostructures opens up the possibility to create an unprecedented architecture using 2D atomic layer building blocks. The thermal transport properties of such mixed heterostructures, critical to various applications in nanoelectronics, however, have not been thoroughly explored. Herein, we construct two configurations of multilayer in-plane graphene/hexagonal boron nitride (Gr/*h*-BN) heterostructures (i.e. mixed heterostructures) *via* weak van der Waals (vdW) interactions and systematically investigate the dependence of their interfacial thermal conductance (ITC) on the number of layers using non-equilibrium molecular dynamics (NEMD) simulations. The computational results show that the ITC of two configurations of multilayer in-plane Gr/*h*-BN heterostructures (MIGHHs) decrease with increasing layer number *n* and both saturate at *n* = 3. And surprisingly, we find that the MIGHH is more advantageous to interfacial thermal transport than the monolayer in-plane Gr/*h*-BN heterostructure, which is in strong contrast to the commonly held notion that the multilayer structures of Gr and *h*-BN suppress the phonon transmission. The underlying physical mechanisms for these puzzling phenomena are probed through the analyses of heat flux, temperature jump, stress concentration factor, overlap of phonon vibrational spectra and phonon participation ratio. In particular, by changing the stacking angle of MIGHH, a higher ITC can be obtained due to the thermal rectification behavior. Furthermore, we find that the ITC in MIGHH can be well-regulated by controlling the coupling strength between layers. Our findings here are of significance for understanding the interfacial thermal transport behaviors of multilayer in-plane Gr/*h*-BN heterostructure, and are expected to attract extensive interest in exploring its new physics and applications.

**Keywords:** mixed heterostructures, vdW interactions, interfacial thermal conductance, phonon transmission, coupling strength

# 1. Introduction

Van der Waals (vdW) heterostructures have attracted enormous interest in the past decade due to their novel device functionalities and superior electronic and thermal regulation properties.[1-8] Most of these captivating properties are strongly dependent on the different spatial configurations of heterostructures, typically including the vertical heterostructures[1, 9-12] that vertically stacked multiple two-dimensional (2D) materials and the in-plane heterostructures[13-17] that seamlessly stitched together two different atomic monolayers through covalent bonds. Due to the weak interactions of vdW force in the vertical heterostructures, the heat flux through them is much lower than that through the in-plane heterostructures,[7, 18] which makes the in-plane heterostructures have a broader application prospect as the thermal management materials. Still, for the vertical heterostructures, they can be considered as the basis materials for designing the next generation of high-performance wearable devices and the vdW heterostructure-enabled sensors, thanks to the diversity of their spatial configurations.[19-21] Since each configuration has outstanding advantages, it is conceivable that the combination of vertical and in-plane heterostructures will make it possible to create an unprecedented architectures by using 2D atomic layer building blocks.[9]

Graphene/hexagonal boron nitride (Gr/*h*-BN) heterostructure, which benefits from the small lattice mismatch between Gr and *h*-BN, has been successfully synthesized both vertical[22-24] and coplanar[2, 25, 26] configurations, paving the way for constructing the mixed heterostructures. In parallel to the efforts on pursuing a more sophisticated synthesis method, the interfacial thermal transport properties of Gr/*h*-BN heterostructure, which play a pivotal role in the high-performance thermal interface materials (TIMs) and heterostructures devices,[27-29] are urgently needed to be understood. Accordingly, using a transient heating technique, Zhang *et al.*[30] showed that the interfacial thermal resistance in the vertical Gr/*h*-BN heterostructure is affected by interatomic bond strength, heat flux direction and functionalization. Relatively, Chen *et al.*[31, 32] found that the in-plane Gr/*h*-BN heterostructure exhibits a negative differential thermal resistance behavior, which is caused by the phonon resonance effect and the lattice vibration mismatch. In addition, topological defects,[7] doping and interface topography optimization[33] were employed to effectively improve the thermal energy transport across the in-plane Gr/*h*-BN heterostructure interface. These previous studies have enriched our understanding of the interfacial thermal transport properties of Gr/*h*-BN heterostructure. However, most of the Gr/*h*-BN heterostructure simulations are limited to a single vertical or in-plane configuration without paying attention on their combinations, which may create many interesting thermal transport properties and will contribute to the design of functional heterostructures with high-density multi-pixel capabilities.

In the present study, the in-plane Gr/*h*-BN heterostructure is stacked vertically *via* weak vdW interactions and two types of multilayer in-plane Gr/*h*-BN heterostructures (MIGHHs), i.e. mixed heterostructures, are constructed. It has been demonstrated that the thermal conductivity of multilayer Gr and *h*-BN decreases with the increasing number of layers,[29, 34-39] which prevents them from creating new applications in various emerging fields. Therefore, it is a meaningful topic to explore whether the change in the number of layers will have the same effect on the thermal transport properties across the mixed heterostructures interface. To this end, using a series of molecular dynamics (MD) simulations, the interfacial thermal conductance (ITC) of two types of MIGHHs are investigated. Additionally, we further study the effect of coupling strength between different in-plane Gr/*h*-BN heterostructure layers on the ITC. To reveal the interfacial thermal transport mechanisms in MIGHH, the interfacial stress distributions, phonon density of states (PDOS) in the frequency domain and phonon participation ratio (PPR) reflecting the phonon characteristics are performed to analyze and compare in our research.

## 2. Models and methods

To model the MIGHH, constructing the in-plane Gr/*h*-BN heterostructure is a top priority. As previous theoretical[40] and experimental[40, 41] studies have shown that the zigzag stitching edges are preferably formed in the in-plane Gr/*h*-BN heterostructure, so only the interface of the zigzag arrangement is considered here. Figure 1a depicts two different structures of in-plane Gr/*h*-BN heterostructure, called C-NB and C-BN, respectively, depending on the bonding type of different atoms at the interface. It is known that although both C-N and C-B interactions are covalent in nature, the strength of C-N bond is higher than that of C-B bond, leading to the advantage of the C-N interface for phonon transmission.[7, 42, 43] Therefore, when constructing the MIGHH model, we use the in-plane Gr/*h*-BN heterostructure of C-NB, while C-BN is only used for reference and comparison.

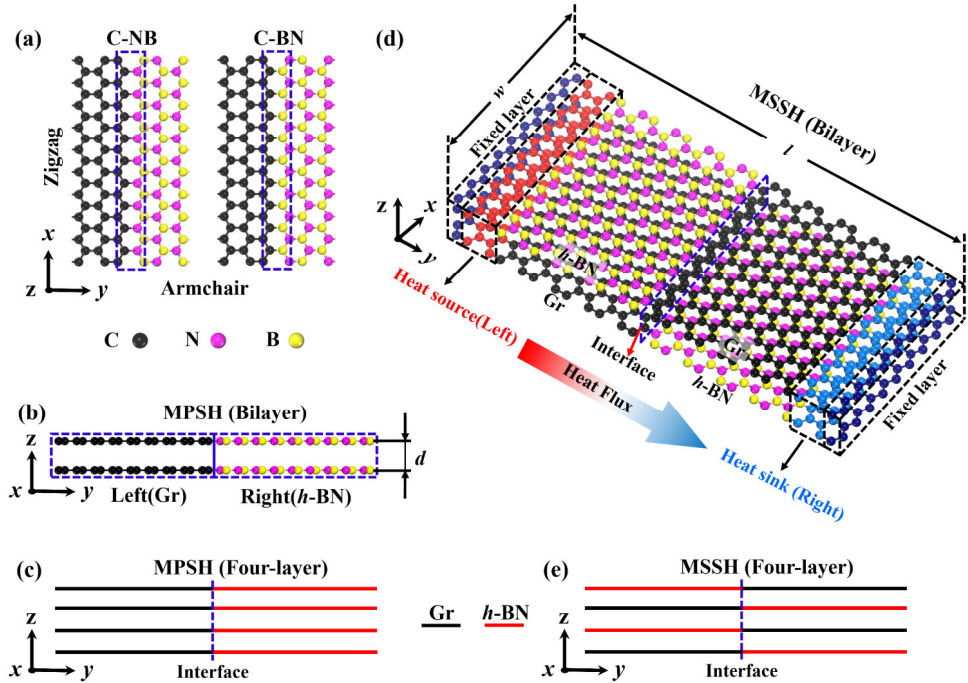

**Fig. 1.** The schematic models of monolayer and multilayer in-plane Gr/*h*-BN heterostructures. (a) Top view for the atomic model of monolayer in-plane Gr/*h*-BN heterostructure including C-NB and C-BN structures. (b) The atomic model of MPSH (bilayer) viewed from the side. (c) Side view for the simplified model of MPSH (four-layer), where black and red lines represent Gr and *h*-BN, respectively. (d) Perspective view for the atomic model of MSSH (bilayer). Heat source and sink are assigned to the both ends of the simulation systems (excluding the fixed layers of 15.5 nm), generating a heat flux across the interface. (e) Side view for the simplified model of MSSH (four-layer).

In this work, two types of MIGHHs are constructed. One is to directly stack the in-plane Gr/*h*-BN heterostructure in order, named as multilayer parallel stacked heterostructure (MPSH), as shown in Fig. b-c. The other is to rotate the even-numbered in-plane Gr/*h*-BN heterostructure by 180 degrees on the basis of the MPSH configuration to form the multilayer staggered stacked heterostructure (MSSH), as shown in Fig. d-e. To construct the initial configuration of the MIGHH, the interlayer distance $d$ (c.f. Fig. 1b) is temporarily set at 0.35 nm, and it will be automatically optimized to a reasonable value after the system reach a steady state, depending on the interactions strength of the vdW force between the layers. Meanwhile, the $y$-axis is set to the direction in which the heat flux is applied, while the $x$-axis is parallel to the heterostructure interface, as schematically shown in Fig. 1d. The dimensions of all the simulated systems are $w = 62.12$ and $l = 215.50$ nm in the $x$-axis (along the zigzag direction) and $y$-axis (along the armchair direction), respectively.

All the MD simulations are performed using the Large-Scale Atomic/Molecular massively Parallel Simulator (LAMMPS) package.[44] Fixed boundary conditions are applied in the transport direction, whereas periodic boundary conditions are applied in the direction of $w$. To allow the interlayer distance to be adjusted freely, free boundary conditions are applied in the transverse direction ($z$-direction). The optimized C-B-N parameters of the Tersoff potential given in ref. (45) are used to describe the atomic covalent interactions between C, N and B atoms, since they have been proved to be of high quality in evaluating thermal transport properties.[7, 31, 46] Correspondingly, the Lennard-Jones (LJ) potential is used to build the weak vdW force between different in-plane Gr/$h$-BN heterostructure layers and is adopted as:

$$V(r) = 4\chi\varepsilon \left[ \left(\frac{\sigma}{r}\right)^{12} - \left(\frac{\sigma}{r}\right)^{6} \right], \tag{1}$$

where $r$ is the distance between two atoms; $\varepsilon$ and $\sigma$ are the energy and distance constants, respectively. The parameter $\chi$ is used to control the coupling strength between different in-plane Gr/$h$-BN heterostructure layers. The LJ parameters are listed in Table S1, which are calculated from a universal force field (UFF) model.[47] The cutoff distance of the LJ potential is set as 12 Å. We have tested the effect of using different cutoff distances on simulation results in Supporting Information (see Table S2).

The ITC of the mixed Gr/$h$-BN heterostructures is calculated by using the non-equilibrium molecular dynamics (NEMD) method, and the equations of atomic motion are integrated with a time step of 0.5 fs, which ensures good energy conservation. Each initial structure is first equilibrated in the constant atom number, volume and temperature ensemble (NVT ensemble) with Nosé-Hoover thermostat[48, 49] at 300 K for

2 ns (4 million time steps). Then, the equilibrated system is switched to an NVE ensemble. Meanwhile, we generate the non-equilibrium heat flux $J$ by applying the Bussi-Donadio-Parrinello (BDP) thermostat[50] (more details and corresponding parameter settings can be found in Supporting Information) to the heat source and sink, with a higher temperature of 330 K and a lower temperature of 270 K, respectively, as schematically shown in Fig. 1d. When steady state is achieved, the heat flux $J$ can be calculated by continuously adding or removing energy in the thermostated regions at an energy exchanged rate $dE/dt$, then can be expressed as:

$$J = \frac{dE/dt}{A}, \qquad (2)$$

where $A$ is the cross-sectional area of the simulation system in the direction perpendicular to heat transport. The thickness of monolayer in-plane Gr/$h$-BN heterostructure is chosen as 0.335 nm and that of MIGHH is the product of 0.335 nm and the layer number. The non-equilibrium heat flux is imposed by using the local thermal baths for 6 ns in total, and all the systems reach a steady state after 3 ns. Therefore, in the production stage, we sample the local temperatures and the accumulated energy exchange (see Fig. 2a) between the system and the thermal baths within the last 3 ns. Finally, the time-averaged temperature jump $\Delta T$ (c.f. Fig. 2b) is obtained by linearly fitting these data, and the ITC $G$ of the steady system can be calculated according to Fourier's law as:

$$G = \frac{J}{\Delta T}. \qquad (3)$$

It is worth mentioning that in order to avoid the interface dislocations of different in-plane Gr/$h$-BN heterostructure layers during the equilibrium process, resulting in multiple temperature jumps when heat flux is applied, we have frozen some of the atoms at both ends of the system (15.5 nm in total) before the equilibrium, as visualized in Fig. 1d. For each simulation system, we performed five independent runs and calculated the error bounds in terms of the standard error.

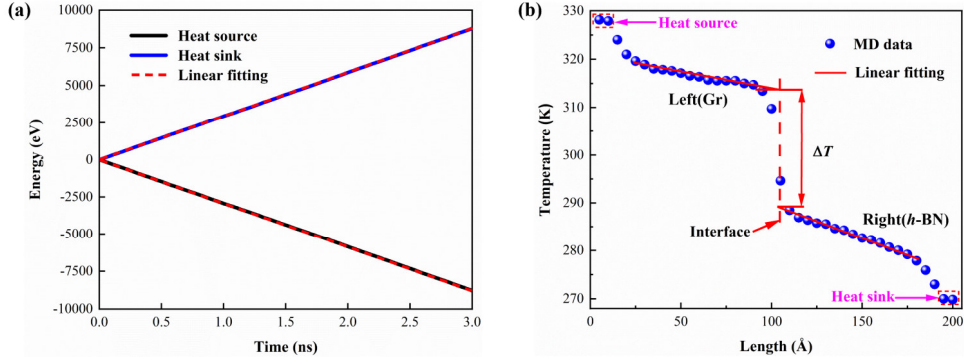

**Fig. 2.** (a) The accumulated energy of the BDP thermostat (averaged over the heat source and sink) as a function of the time in steady state. The energy exchanged rate $dE/dt$ is calculated as the slope of the linear fit (red dashed lines). (b) A typical steady-state temperature profile for the mixed heterostructure system (MPSH, bilayer). The temperature can be fitted by a linear function on each side of the interface of the simulation system, excluding the nonlinear region around the heat source or sink, and then the time-averaged temperature jump $\Delta T$ can be extracted as the difference between the two linear functions at the interface. The number of blocks of heat source and sink is set as 2.

The PDOS is a simple method and powerful tool for characterizing phonon activities in materials,[51] and it is computed from the Fourier transform[52] of the velocity autocorrelation function (VACF) of all the atoms as:

$$\text{PDOS}(\omega) = \int_{-\infty}^{\infty} e^{i\omega t}\, \text{VACF}(t)dt, \tag{4}$$

where $\text{PDOS}(\omega)$ is the total PDOS at the vibrational frequency $\omega$ and $\text{VACF}(t)$ is given by:

$$\text{VACF}(t) = \frac{1}{N}\sum_{i=1}^{N}\langle v_i(0)v_i(t)\rangle. \tag{5}$$

$v_i(t)$ is the velocity vector for particle $i$ at time $t$, $N$ is the number of atoms in the system and the space-average is the average over the particles. The ensemble average (denoted by $\langle\rangle$) in Eqn. 5 is realized by time-averaging over a period of 15 ps, with the sample velocities extracted from the simulation every 5 fs. Since VACF is normalized and dimensionless, the unit of PDOS is 1/THz. For polarized PDOS, the correlation function is calculated as $\langle v(x)v(x) + v(y)v(y)\rangle$ and $\langle v(z)v(z)\rangle$ for the in-plane and out-of-plane PDOS, respectively. To quantitatively the PDOS mismatch of MIGHH on

the two side of the interface, the overlap area factor $S$ is defined as:

$$S = \int_0^\infty \min\{P_{\text{Gr}}(\omega), P_{h-\text{BN}}(\omega)\}\,d\omega, \qquad (6)$$

where $P_{\text{Gr}}$ and $P_{h-\text{BN}}$ are the two PDOS of Gr and $h$-BN, respectively.

Calculating the phonon participation ratio is another effective way to provide insight on the phonon activities, especially for quantitatively describing the phonon localization effect. Without lattice dynamic calculation, the phonon participation ratio at any temperature can be calculated directly from the MD simulations, thereby implicitly including the all-order of anharmonic scatterings,[51] and is defined as:[53]

$$\text{PPR}(\omega) = \frac{1}{N}\frac{(\sum_i \text{PDOS}_i(\omega)^2)^2}{\sum_i \text{PDOS}_i(\omega)^4}, \qquad (7)$$

where $N$ is the total number of atoms, and $\text{PDOS}_i(\omega)$ is the single-particle phonon density of states obtained by calculating a single-particle VACF.

## 3. Results and discussion

**3.1. ITC of multilayer in-plane Gr/$h$-BN heterostructure.** Firstly, we focus on the layer effect on the interfacial thermal conductance of the MIGHH by varying the layer number $n$ from 2 to 6. From Fig. 2 and Eqn. 3, we obtained the values of ITC of different heterostructure systems at room temperature (300 K), which are shown in Fig. 3a. It is seen that the ITC of C-NB is larger than the C-BN, with the values of 4.35 $\pm$ 0.12 and 3.54 $\pm$ 0.07 GW m$^{-2}$ K$^{-1}$, respectively, which are consistent well with some previous studies.[7, 54, 55] However, these values are about two to three times larger than those of other studies,[33, 56-58] and even larger. The discrepancy stems from the differences in simulation dimensions, boundary conditions, bonding types, especially the thermostatting methods that generate heat flux. As can be clearly seen from Fig. 3a that the ITC of MSSH and MPSH decrease with increasing layer number $n$ and both saturate at $n = 3$. Nevertheless, both the saturation values are still slightly larger than the ITC value of monolayer in-plane Gr/$h$-BN heterostructure, indicating that the MIGHH is more conducive to interfacial thermal transport. As mentioned above, the multilayer structures of Gr and $h$-BN suppress the phonon thermal transport, since the out-of-plane flexural phonon modes are restricted by the interlayer vdW interactions.[36-38] Therefore, the behavior of MIGHH to enhance the ITC is very confusing, and we will analyze its mechanisms later.

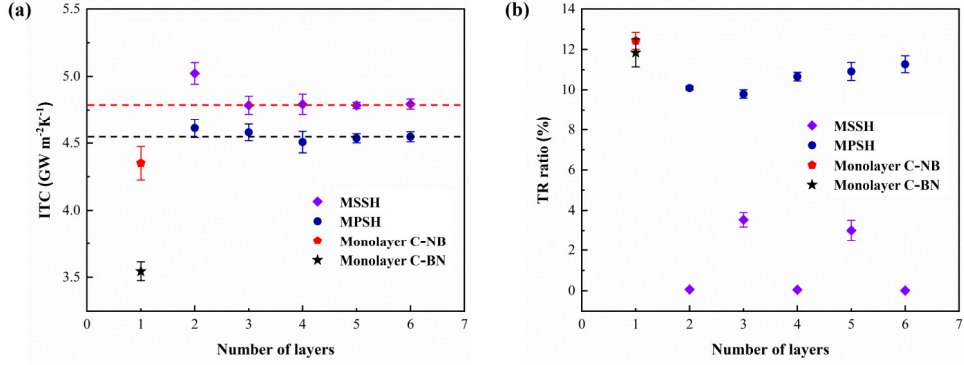

**Fig. 3.** (a) ITC and (b) TR ratio of different heterostructure systems as a function of the number of layers. The red and black dotted lines in (a) are guide for the eyes.

Another striking feature observed here is that the ITC value saturated by MSSH is larger than that saturated by MPSH, suggesting that better thermal transport effect can be obtained by changing the stacking angle of MPSH. This result is foreseeable, because it has been proved in many studies[31, 46, 59] that the magnitude of heat flux from h-BN to Gr domain is higher than that in the opposite direction, indicating a clear thermal rectification (TR) performance in the monolayer in-plane Gr/h-BN heterostructure. In support of this, we further calculate the TR ratio in different heterostructure systems via:

$$\text{TR}(\%) = \frac{J_H - J_L}{J_L} \times 100\%, \tag{8}$$

where $J_H$ and $J_L$ are the high and low heat currents obtained by swapping the temperature bias of the system. And when the temperature bias is inverted, the temperature difference between thermal reservoirs remains the same. Here, the $J_H$ and $J_L$ are the magnitudes of the heat flux in the directions from the right to left region and the left to right region (see Fig. 1b and d), respectively.

Figure 3b illustrates the TR ratios for the different heterostructure systems. Indeed, in the monolayer in-plane Gr/h-BN heterostructure, heat flows preferentially from the h-BN to Gr domain, which demonstrates pronounced TR behavior. Hence, the reason for contributing to this striking feature is as follow: due to the difference in the structural arrangement between MPSH and MSSH (see Figs. 1b and 1d, respectively), when the heat flux direction is from left to right, the TR behavior leads to a larger heat flux in the MSSH configuration, further leading to the improvement of ITC. Besides, we find that the TR of MSSH changes non-monotonically (inverted V-shaped) with the increase of $n$, and the thermal rectification effect disappears (i.e. TR ratio can be considered statistically zero) when $n$ is even. This is because the structures of the MSSH are exactly the same at both ends at $n$ = even. Moreover, at $n$ = odd (excluding $n$ = 1), the TR of MSSH is much smaller than that of MPSH, indicating that it is difficult to achieve both high ITC and excellent TR performance in MIGHH at the same time. Note that the heat flux used to calculate the ITC in this study is from left to right, and the TR ratio is calculated only in this subsection.

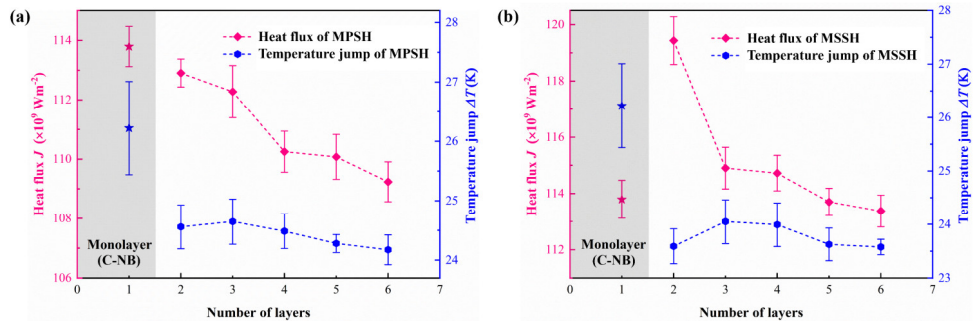

**Fig. 4.** Heat flux and temperature jumps of (a) MPSH and (b) MSSH as a function of the number of layers.

Below, we are going to deeply explore the physical origin of MIGHH improving ITC. According to Eqn. 3, the ITC value is determined by the heat flux $J$ and temperature jump $\Delta T$, so we collected the average values of $J$ and $\Delta T$ in different heterostructure systems, as shown in Fig. 4. At this point, the focus of the problem shifts to exploring the root causes of changes in these two physical quantities. From Fig. 4a, we can see at a glance that both $J$ and $\Delta T$ of MPSH decrease monotonically with the increase of $n$, and both are smaller than those of monolayer in-plane Gr/$h$-BN heterostructure. For the $\Delta T$ of MSSH (see Fig. 4b), however, it increases slightly and then decreases as the number of layers increases, while the $J$ of MSSH surges at $n$ = 2 (due to the influence of TR behavior), and then also decreases monotonously with the increase of $n$. The behavior that the $J$ and $\Delta T$ of MPSH and MSSH eventually decrease with the increase of *n* further confirms the results in Fig. 3a and is expected to result from the mechanisms of stress distributions, phonon coupling, and phonon localization near the interfaces.

The interfacial thermal transport properties can be characterized laterally by the stress field, which has been demonstrated in many studies.[7, 54, 56, 60] Accordingly, we calculate the initial stress distribution in different heterostructure systems after they are fully relaxed. The von Mises stress on each atom is calculated as:

$$\sigma_m = \left\{\frac{1}{2}[(\sigma_{11} - \sigma_{22})^2 + (\sigma_{11} - \sigma_{33})^2 + (\sigma_{33} - \sigma_{11})^2 + 6(\sigma_{12}^2 + \sigma_{23}^2 + \sigma_{31}^2)]\right\}^{\frac{1}{2}} \tag{9}$$

where $\sigma_{11}$, $\sigma_{22}$, and $\sigma_{33}$ are the normal stress along the *x*, *y*, and *z* directions, respectively; $\sigma_{12}$, $\sigma_{23}$, and $\sigma_{31}$ are the shear stress. From Fig. 5, a clear stress concentration is observed near the interface in different heterostructure systems due to the lattice mismatch between Gr and *h*-BN and is more severe in the Gr domain due to its higher in-plane stiffness.[61] What is more, the average stress near the interface of C-NB is lower than that of C-BN, and their corresponding temperature jumps are 26.22 ± 0.79 and 29.88 ± 0.33 K, respectively. The mitigating/severe interfacial stress level of the C-NB/C-BN intensifies/constrains vibrations of atoms near the interface, thus leading to a lower/higher $\Delta T$. Unfortunately, when comparing the monolayer in-plane Gr/*h*-BN heterostructure with the MPSH (bilayer) (Figs. 5a and 5c, respectively), we find that the difference in stress distribution near the interface between the two is not obvious enough. Thus, to further quantitatively analyze the contribution of the interfacial stress field, the stress concentration factor $V_s$ is calculated by:

$$V_s = \frac{\sigma_A}{(\sigma_B + \sigma_C)/2}, \tag{10}$$

where $\sigma_A$ and $\sigma_B$, $\sigma_C$ are the averaged stress extracted from the stress concentration field near the interface (Region A in Fig.5a) and far from the stress concentration field (Region B and C in Fig.5a), respectively.

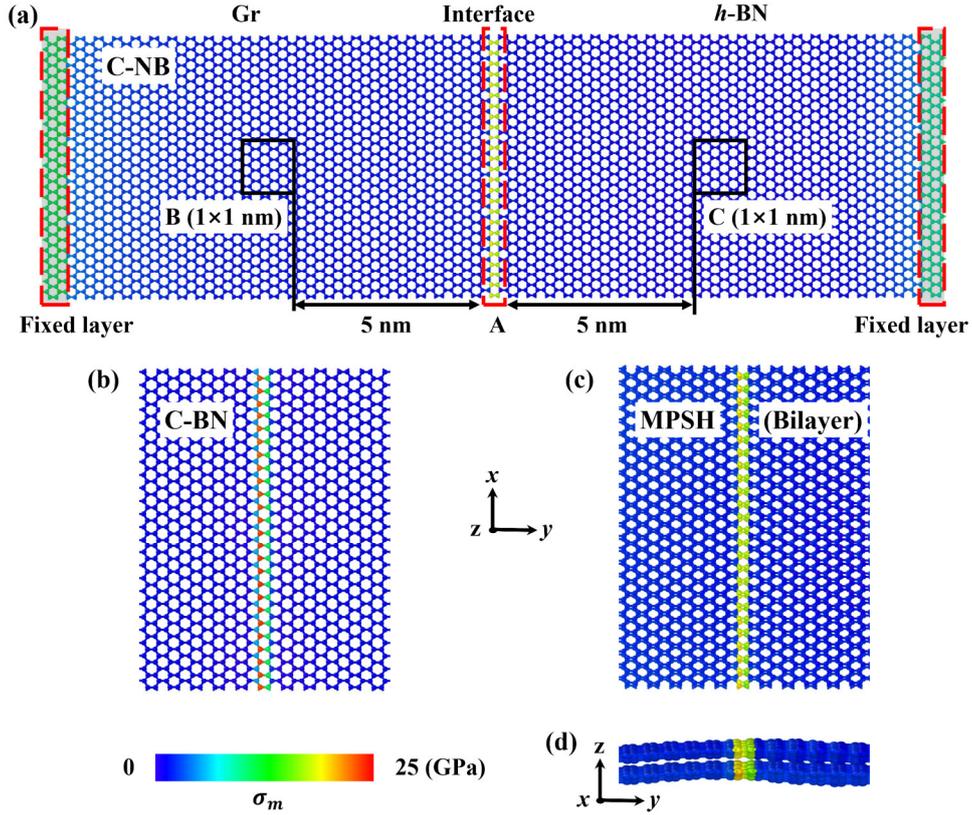

**Fig. 5.** Snapshots of von Mises stress distribution in different heterostructure systems. (a) Von Mises stress distribution in the whole configuration of C-NB. The average stress for the atoms in fixed layers is relatively high, since these atoms have not been relaxed. (b) Von Mises stress distribution in the C-BN near the interface. The snapshot is representative section of C-BN, since the whole configuration is too long to be viewed in details. (c) Top and (d) side views for von Mises stress distribution in MPSH (bilayer).

Figure 6a intuitively shows the numerical trend of stress distribution in MPSH (bilayer) along the *y*-direction. Meanwhile, in Fig 6b, the variation trends of $V_s$ for MSSH and MPSH with the number of layers are consistent well with that of temperature jumps in Fig. 4. Furthermore, the $V_s$ of C-NB is lower than that of C-BN. These self-consistent results suggest that the reduction in the stress concentration factor $V_s$, caused by the effect of vdW force to intensify the vibration of atoms near the interface, is responsible for the decline in temperature jump $\Delta T$ of MIGHH. In addition, we find that the $V_s$ and $\Delta T$ of MIGHH dramatically decrease at $n = 2$. The possible reason for this anomaly is that the atomic vibration near the interface responds rapidly to the interactions of vdW force at the initial stage, but the response slows down as the

$n$ increases. Therefore, with the increase of the number of layers ($n > 2$), the declining trends of $V_s$ and $\Delta T$ of MIGHH slow down or even increase slightly, but both eventually show a downward trend, as shown in Fig. 4 and 6b.

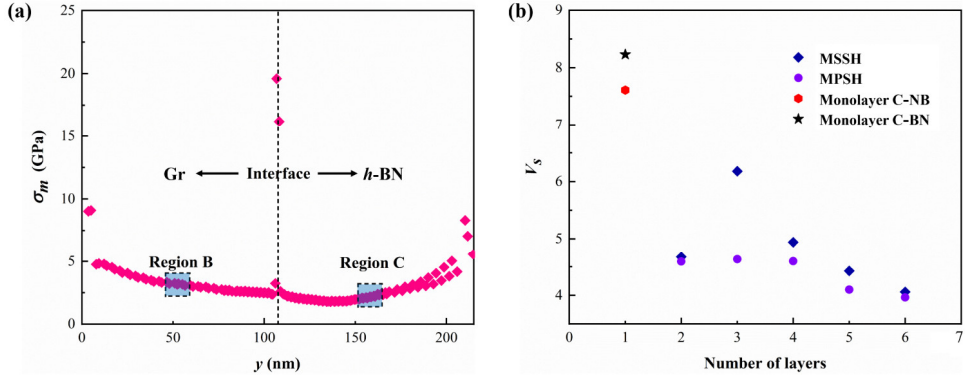

**Fig. 6.** (a) The variation of von Mises stress at different locations in MPSH (bilayer) along the $y$-direction. (b) The dependence of stress concentration factor $V_s$ on the number of layers in different heterostructure systems.

Next, we explore the internal mechanisms for the monotonous decrease of heat flux of MIGHH with the increase of $n$ by understanding the phonon activities in depth. According to Eqn. 4-6, the popular PDOS and its overlap representing the phonon coupling degree are obtained by changing the speed random number and performing three independent calculations. Due to the complexity of the MSSH configuration, it is difficult to separate the PDOS of Gr and $h$-BN, so we only calculate the PDOS of the MPSH as a typical representative. Figure 7a-c clearly show that the out-of-plane PDOS of Gr and $h$-BN in MPSH have different degrees of decline in the low frequency region (0-10 THz) comparing with that in monolayer in-plane Gr/$h$-BN heterostructure. And as the number of layers increases, the decreasing region increases, which directly leads to the reduction in out-of-plane overlap $S_o$ (see in Fig. 7e). Actually, in MIGHH, the out-of-plane flexural phonon modes of Gr and $h$-BN are restricted by the interlayer vdW interactions,[36-38] causing the reduced PDOS of Gr and $h$-BN in the low frequency region, thus resulting in the reduction of $S_o$. Therefore, the weakening of out-of-plane phonon coupling weakens the phonon thermal transport ability across the interfaces, resulting in the decrease of $J$ in MIGHH.

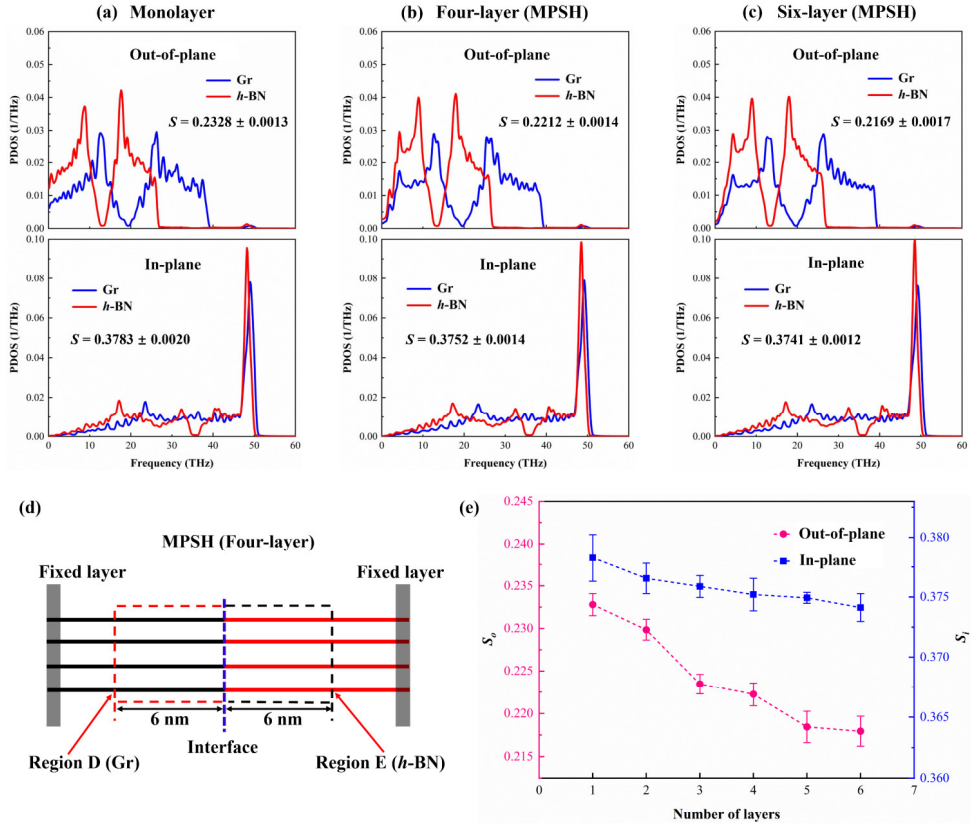

**Fig. 7.** (a)-(c) Out-of-plane and in-plane PDOS in different heterostructure systems. (d) A schematic diagram of the regions selected to calculate PDOS. To avoid the interference from the fixed layers, we chose the regions on both sides of the interface and away from the fixed layers. (e) The dependences of overlap of out-of-plane ($S_o$) and in-plane ($S_i$) PDOS on the number of layers in different heterostructure systems.

In comparison with the out-of-plane PDOS resulting in a large drop in overlap, the in-plane PDOS, which has been shown in many studies[32, 46, 62, 63] to play a minor role in phonon coupling, is not many differences as seen in Fig. 7a-c. Although in Fig. 7e, the overlap ($S_i$) of in-plane PDOS in different heterostructure systems decreases monotonically as the $n$ increases, the decrease amplitude is much smaller than that of $S_o$. This proves that in MIGHH, the out-of-plane acoustic phonons play a dominant role in the thermal transport across such interface. Therefore, from the perspective of phonon coupling theory, the reduction in the heat flux of MIGHH is due to the weakening of phonon coupling between Gr and $h$-BN, and the weakening of out-of-plane phonon coupling is the main responsibility.

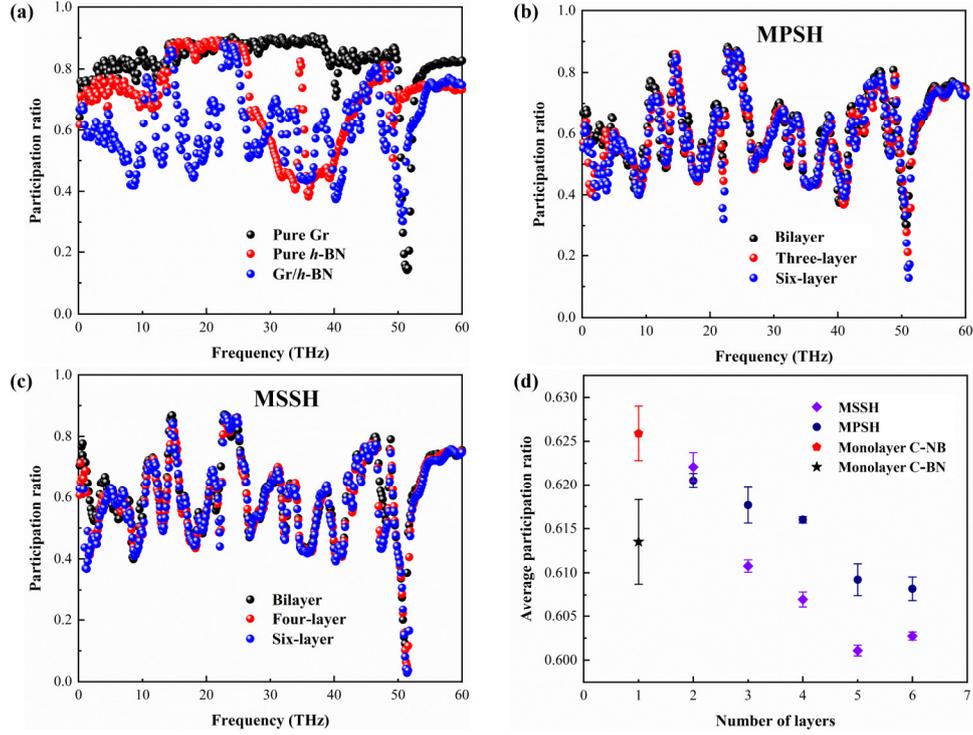

**Fig. 8.** (a) Phonon participation ratio of monolayer Gr/*h*-BN heterostructure and pure Gr and *h*-BN configurations. (b) and (c) Comparison of phonon participation ratio in MIGHH. (d) The average participation ratio in different heterostructure systems as a function of the number of layers. Similar to the PDOS calculation, the average phonon participation ratio is obtained by performing three independent calculations.

To further capture the phonon activities (especially for the degree of phonon localization) and their effects on the reduction of heat flux in MIGHH, we calculate the PPR using Eqn. 7. It measures the fraction of atoms participating in a given mode and can effectively indicate the localized modes with $O$ ($1/N$) and delocalized modes with $O$ (1).[64] Therefore, the PPR is widely used to reveal the phonon confinement and thermal rectification in different Gr/*h*-BN heterostructure systems.[31, 46, 59] In our analyses, PPR < 0.4 is taken as the criterion to characterize the localized modes. Figure 8a shows an overall reduced PPR in monolayer Gr/*h*-BN heterostructure (compared with the pure Gr and *h*-BN), due to the mismatch of the intrinsic phonon modes of Gr and *h*-BN. In spite of this, most of the phonon modes in the monolayer Gr/*h*-BN heterostructure have PPR greater than 0.4, showing a characteristic of delocalized mode. However, from Fig 8b-d, we can see that the PPR of MPSH and MSSH decreases with the increase of *n*, whether in low-frequency phonons (0-10 and 20-25 THz) or high-frequency phonons (50-60 THz). In other words, as *n* increases, more and more phonon modes in MPSH and MSSH have PPR less than 0.4, showing a characteristic of localized mode. In fact, the contribution to thermal transport mainly comes from the delocalized modes rather than the localized modes. Therefore, these analyses show that the increase in the number of layers induces a greater degree of phonon localization in MIGHH, causing phonon thermal transport to be blocked, eventually leading to a decrease in heat flux.

To get a better physical picture about the phonon localization, we further provide the information about the spatial distribution of a specific mode ($\Lambda \in$ PPR < 0.4) by[65]

$$\phi_{i\alpha,\Lambda} = \frac{\int_\Lambda PDOS_{i\alpha} d\omega}{\frac{1}{N} \sum_{1 \leq i \leq N} \int_\Lambda PDOS_{i\alpha} d\omega}. \tag{11}$$

A larger value of $\phi_{i\alpha,\Lambda}$ indicates stronger localization of modes $\Lambda$ on the *i*th atom. The in-plane and out-of-plane phonon modes ($\alpha = x, y$ and $z$) are considered here. Figure 9 shows the spatial distribution of the phonon localized modes near the interface in different heterostructure systems at 300 K. Obviously, the intensity of the phonon localized modes on the Gr domain is greater than that on the *h*-BN domain. That is to say, the Gr domain is the main bottleneck of phonon thermal transport channels, which has been confirmed in the studies of thermal rectification.[31, 46] Moreover, it is observed from Fig. 9 that with the increase of *n*, the intensity of the phonon localized modes in MPSH and MSSH gradually increases, i.e. more phonon modes are converted from delocalized to localized modes both in Gr and *h*-BN domains, thus causing a gradual blockage of phonon transmission channels. When comparing MPSH and MSSH with the same number of layers (see a3 and b3 in Fig. 9), it can be found that phonon localization is more severe in MSSH including Gr and *h*-BN domains, which may be attributed to the disrupted symmetry along the *z* direction in the MSSH configuration.

Stated thus, based on the analyses of two physical quantities, heat flux *J* and temperature jump $\Delta T$, we can summarize as follows: (1) Owing to the effect of vdW force, the interface stress concentration $V_s$ decreases with the increase of *n*, leading to the decrease of $\Delta T$ in MIGHH. (2) With the weakening of phonon coupling between Gr and *h*-BN and the increase of phonon localized modes, the heat flux decreases with the increase of $n$. (3) Due to the combined effect of (1) and (2), the ITC of MIGHH eventually decreases with the increase of $n$, saturates at $n = 3$, and improves.

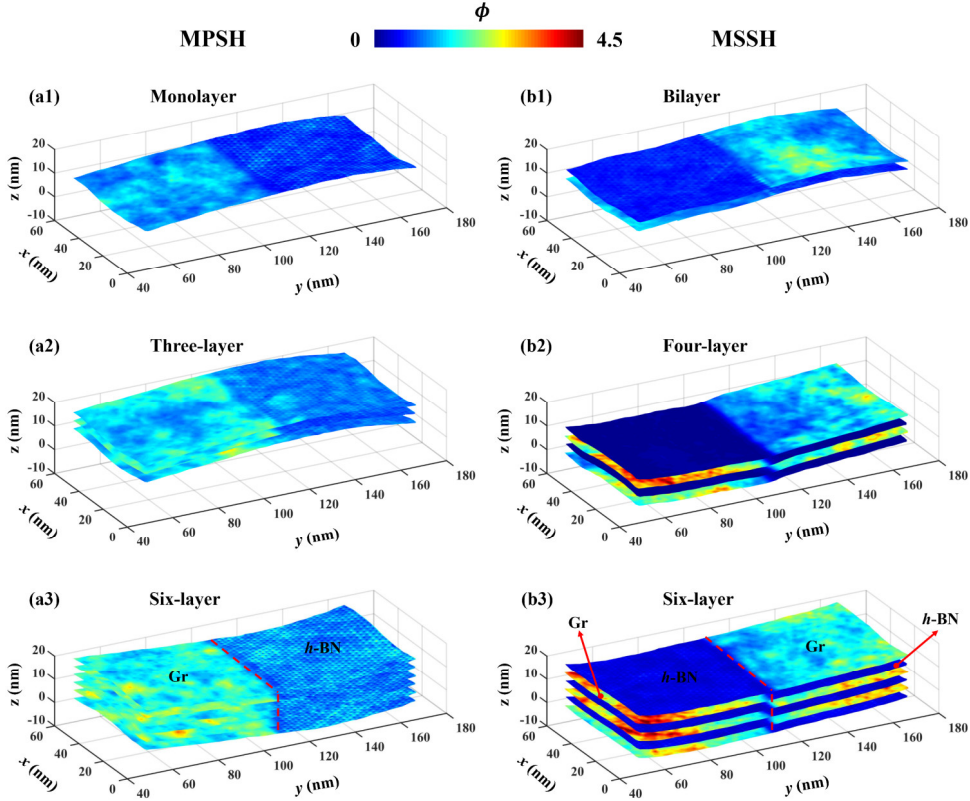

**Fig. 9.** The typical spatial distribution of the phonon localized modes in different heterostructure systems. The intensity of the localization is depicted according to the color bar.

**3.2. Effect of the coupling strength between layers.** We know that in the vertical heterostructures, the vdW force coupling strength between layers can greatly affect the thermal transport properties, which has been revealed in many studies.[11, 58, 62] Coincidentally, in MIGHH, the atoms vibration near the interface is stimulated by the vdW force, so it is necessary to explore the influence of coupling strength on the ITC of MIGHH. To this end, we adjust the coupling strength by changing the parameter χ from 2 to 10. For convenience, here we only consider the effect of different coupling strength on the ITC of MPSH (bilayer), and the conclusions obtained are also true in the MSSH configuration.

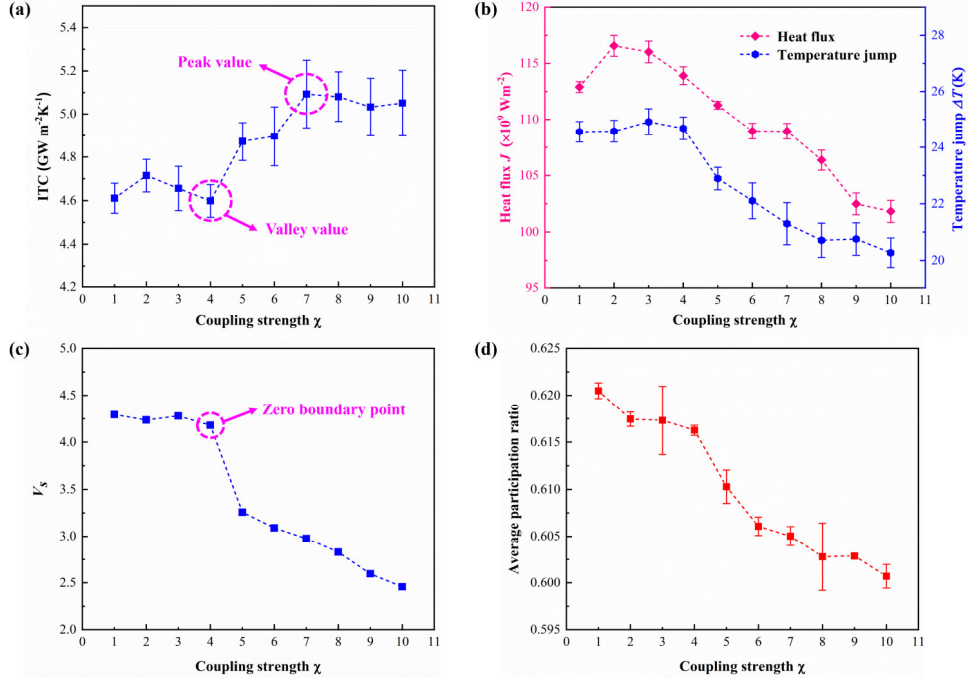

**Fig. 10.** (a) The ITC, (b) heat flux $J$ and temperature jump $\Delta T$, (c) stress concentration $V_s$ and (d) average participation ratio of MPSH (bilayer) as a function of the coupling strength $\chi$.

From Fig. 10a, it can be seen that as the coupling strength $\chi$ increases, the ITC of MPSH (bilayer) first increases then drops to the valley at $\chi = 4$, and then rises to the peak and converges at $\chi = 7$. That is to say, the ITC of MPSH (also for MSSH) can be either enhanced or suppressed by controlling coupling strength $\chi$. To gain insights into the underlying mechanisms for this phenomenon, the heat flux $J$, temperature jump $\Delta T$, stress concentration factor $V_s$, overlap of PDOS and phonon participation ratio analyses are also carried out here. Fig. 10b shows that the $\Delta T$ remains unchanged first and then decreases gradually with the increase of $\chi$, while the $J$ increases first and then decreases monotonously with the increase of $\chi$. As stated above, the change of $\Delta T$ is attributed to the change of $V_s$, which is also confirmed by the mutual agreement of Fig. 10b-c. Interestingly, there is a zero boundary point (i.e. $\chi = 4$) for the change of $V_s$, indicating that when $\chi > 4$, the atomic vibration near the interface is sharply intensified, which may provide new inspiration for the regulation of ITC. In addition, we calculate the overlap of in-plane and out-of-plane PDOS in MSPH (bilayer), as shown in Fig. S1. A counter-intuitive phenomenon, in which the overlap of in-plane PDOS increases with the increase of $\chi$, has been observed. This reveals the mechanism for the increase in heat flux at the initial stage of $\chi$ increase. Then, with the increase of $\chi$, the decline (see

Fig. S1) in the overlap of out-of-plane PDOS is dominant, causing the heat flux to eventually decline. Furthermore, the decrease (see Fig. 10d) of phonon average participation ratio also results in the decrease of heat flux. The detailed information on the spatial distribution of the phonon localized modes in MPSH (bilayer) is also presented here, as illustrated in Fig. S2. It can be clearly seen that as the χ increases, the phonon localization in the *h*-BN domain is progressively more severe than that in Gr domain. Hence, we can conclude that the effect of enhanced coupling strength χ on phonon modes in *h*-BN is more significant.

## 4. Conclusions

To summarize, in this paper, we construct two types of multilayer in-plane Gr/*h*-BN heterostructures (including MPSH and MSSH configurations) and systematically investigate the effect of the number of layers on their ITC by performing NEMD simulations. It is interesting to find that the ITC of MPSH and MSSH decrease with increasing layer number *n* and both saturate at *n* = 3. Nevertheless, both the saturation values are still larger than the ITC value of monolayer in-plane Gr/*h*-BN heterostructure, suggesting that the MIGHH is more conducive to interfacial thermal transport. Based on the analyses of heat flux, temperature jump, stress concentration factor, overlap of PDOS and phonon participation ratio, we conclude that the drop in temperature jump originates from the mitigation of stress concentration; the decrease of heat flux results from the weakening of phonon coupling between Gr and *h*-BN and the increase of phonon localized modes, and ultimately, the combined effect of the two leads to these phenomena. Moreover, by comparing the ITC in MPSH and MSSH, it is found that a higher ITC can be obtained by changing the stacking angle of MPSH, due to the thermal rectification behavior. Finally, we explore the effect of coupling strength on the ITC of MPSH (bilayer) and find that the coupling strength can be effective in modulating the ITC in MIGHH. We anticipate that our findings not only can shed some light on exploring the thermal transport properties of other in-plane heterostructures, such as phosphorene/graphene,[15] $MoS_2$/graphene[16] and silicene/graphene,[17] but also provide a new design idea for the applications of multilayer in-plane Gr/*h*-BN heterostructure in thermal management and thermoelectric devices.

## Acknowledgments

The author(s) disclosed receipt of the following financial support for the research, authorship, and/or publication of this article: The authors acknowledge the financial support provided by National Natural Science Foundation of China (Project No. 51506033), Guangxi Natural Science Foundation (Grant No. 2017JJA160108), and Guangxi Colleges and Universities Program of Innovative Research Team and Outstanding Talent.

# Supporting Information for

## "Van der Waals stacked multilayer in-plane graphene/hexagonal boron nitride heterostructure: its interfacial thermal transport properties"

Ting Liang, Ping Zhang[*], Peng Yuan and Man Zhou

School of Mechanical and Electrical Engineering, Guilin University of Electronic Technology

No. 1 Jinji Road, Guilin, Guangxi 541004, China

*corresponding author, E-mail: *pingzhang@guet.edu.cn*

In our simulations, the Bussi-Donadio-Parrinello (BDP) thermostat[1] is adopted for generating heat flux at both ends of the systems. As a reformulation of the Berendsen thermostat,[2] the BDP incorporates a proper randomness into the velocity re-scaling factor $\alpha$ and generates a true NVT ensemble. The velocities are scaled in the following way:

$$\vec{v}_i^{\,scaled} = \alpha \vec{v}_i;$$

$$\alpha^2 = e^{-\Delta t/\tau} + \frac{T_0}{TN_f}(1-e^{-\Delta t/\tau})\left(R_1^2 + \sum_{i=2}^{N_f} R_i^2\right) + 2e^{-\Delta t/2\tau}R_1\sqrt{\frac{T_0}{TN_f}(1-e^{-\Delta t/\tau})},$$

where $\vec{v}_i$ is the velocity of atom *i* before the re-scaling, $T_0$ is the target temperature, $T$ is the instant temperature, $N_f$ is the number of degrees of freedom in the thermostatted system and $\{R_i\}_{i=1}^{N_f}$ are $N_f$ Gaussian distributed random numbers with zero mean and unit variance. Particularly, in the BDP thermostat, $\tau$ is the relaxation time, a free parameter that is crucial to the dynamic properties of the simulations. Chen *et al*.[3] have revealed that different $\tau$ in the Langevin[4] and Nosé-Hoover[5, 6] thermostat would make a big difference in the calculated thermal transport results. Differently, for the BDP thermostat, the dynamic properties are slightly affected by different $\tau$.[7, 1] So in our simulations, the BDP thermostat is chosen to generate heat flux, and the relaxation time $\tau$ is set to 0.01 ps. We have tested that when the $\tau$ changes around 0.01 ps, the obtained interfacial thermal conductance values do not fluctuate greatly.

**Table S1.** LJ potential parameters for carbon (C), nitrogen (N) and boron (B) atoms between different in-plane Gr/*h*-BN heterostructure layers.

| Atom 1 | Atom 2 | Energy constant $\varepsilon$ (meV) | Distance constant $\sigma$ (Å) |
|--------|--------|-------------------------------------|--------------------------------|
| C      | C      | 4.55                                | 3.431                          |
| N      | N      | 2.99                                | 3.261                          |
| B      | B      | 7.81                                | 3.638                          |
| C      | N      | 3.69                                | 3.346                          |
| C      | B      | 5.96                                | 3.534                          |
| N      | B      | 4.83                                | 3.449                          |

**Table S2.** The ITC and TR ratio in MPSH (bilayer) with different cutoff distances in LJ potential. It can be seen from the table that within the error bounds, the ITC and TR values of MPSH are almost identical when different LJ cutoff distances are used, since the van der Waals interactions are very weak relative to the covalent bond. Hence, the conclusions of our work are not affected by the choice of cutoff distance.

| Cutoff distance in LJ (Å) | ITC (GW m$^{-2}$ K$^{-1}$) | TR ratio (%) |
| --- | --- | --- |
| 9 | 4.56 ± 0.09 | 10.39 ± 0.57 |
| 10 | 4.58 ± 0.06 | 9.50 ± 0.62 |
| 11 | 4.53 ± 0.05 | 10.51 ± 0.47 |
| 12 | 4.61 ± 0.07 | 10.10 ± 0.10 |

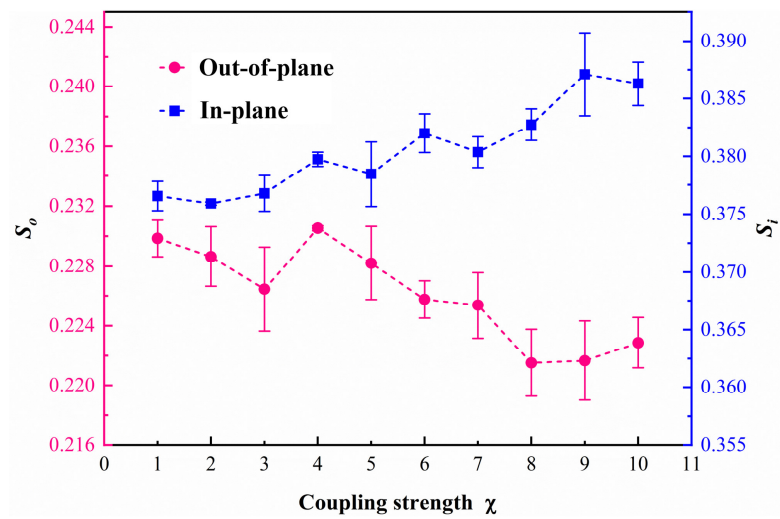

**Fig. S1.** Overlap of out-of-plane ($S_o$) and in-plane ($S_i$) PDOS for MPSH (bilayer) as a function of the coupling strength χ.

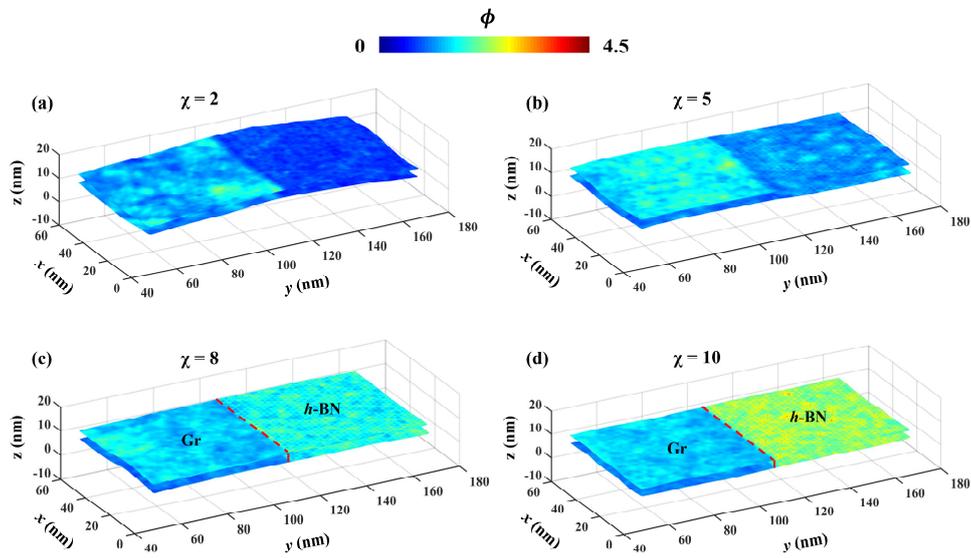

**Fig. S2.** The typical spatial distribution of the phonon localized modes ($\Lambda \in$ PPR < 0.4) in MPSH (bilayer) at different coupling strength χ.